\documentclass{ipart}

\usepackage{amsthm}
\usepackage{amssymb}
\usepackage{amsmath}
\usepackage{mathrsfs}
\usepackage[numbers,square]{natbib}
\usepackage[usenames,dvipsnames]{xcolor}
\usepackage{adjustbox}
\usepackage[pagebackref=false,colorlinks,linkcolor=blue,citecolor=magenta,bookmarksnumbered=true]{hyperref}
\usepackage{lscape,slashed}
\usepackage{caption}
\usepackage{subcaption}
\usepackage{multicol}
\usepackage{graphicx}
\usepackage{subcaption}
\usepackage{tikz}
\usepackage[toc]{appendix}

\startlocaldefs
\theoremstyle{plain}

\definecolor{brown(web)}{rgb}{0.65, 0.16, 0.16}

\endlocaldefs

\begin{document}

\begin{frontmatter}

\title{Testing the Zonal Stationarity of Spatial Point Processes: 
Applied to prostate tissues and trees locations}

\runtitle{Testing the Zonal Stationarity of Spatial Point Processes}
\begin{aug}
 \author{\fnms{Azam} \snm{Saadatjouy} 
 \ead[label=e1]{ a\_saadatjouy@sbu.ac.ir}}
 \address{Department of Statistics,\\ Shahid Beheshti University, G.C.\\Evin, 1983969411, 
 Tehran, Iran.\\
 \printead{e1}} 
 \affiliation{Shahid Beheshti University, G.C.}
 \and
\author{\fnms{Ali Reza} \snm{Taheriyoun}\thanksref{t1}\ead[label=e2]{a\_taheriyoun@sbu.ac.ir}}
\address{ Department of Statistics,\\ Shahid Beheshti University, G.C.\\Evin, 1983969411, 
 Tehran, Iran.\\
\printead{e2}}
\affiliation{Shahid Beheshti University, G.C.}
\thankstext{t1}{Corresponding author}
\and
\author{\fnms{Mohammad Q.} \snm{Vahidi-Asl} 
 \ead[label=e3]{ m-vahidi@sbu.ac.ir}}
 \address{Department of Statistics,\\ Shahid Beheshti University, G.C.\\Evin, 1983969411, 
 Tehran, Iran.\\
 \printead{e3}}
 \affiliation{Shahid Beheshti University, G.C.}
 \runauthor{Saadatjouy, et al.} 

\end{aug}

\received{\sday{22} \smonth{1} \syear{2017}}

\begin{abstract}
We consider the problem of testing the stationarity and isotropy of a spatial point pattern based on the concept of local spectra. Using a logarithmic transformation, the mechanism of the proposed test is approximately identical to a simple two factor analysis of variance procedure when the variance of residuals is known. This procedure is also used for testing the stationarity in neighborhood of a particular point of the window of observation. The same idea is used in post-hoc tests to cluster the point pattern into stationary and nonstationary sub-windows. The performance of the proposed method is examined via a simulation study and applied in a practical data.
\end{abstract}





\end{frontmatter}

\section{Introduction}\label{intro}
\subsection{General aspects}
A point pattern as a realization of a point process is a set of points that are distributed irregularly in a window. The analysis of a point pattern provides information on the geometrical structures formed by the aggregation and interaction between points. The aggregation and interaction of points are characterized by intensity and covariance density functions, respectively. The spectral density function of a stationary point process is the Fourier transform of the complete covariance density function and is estimated in an asymptotically unbiased manner by the spatial periodogram. For a stationary point process all the second-order characteristics (e.g., the complete covariance density function) are only a function of inter-points distances and thus the spectral density function does not depend on the locations of points. So, the structure of the complete covariance density and spectral density functions may vary with location when the stationarity assumption is violated.

Several efficient methods based on the spectral density are available in the literature for studying the spatial structure of stationary processes. The spectral methods were used by \cite{guyon} to investigate the asymptotic properties of several estimation procedures for a stationary process observed on a $d$-dimensional lattice. The approximated locations of events in a spatial point pattern were presented using the intersections of a fine lattice on the observation window by \cite{renshaw1} and the point spectrum was approximated by the lattice spectrum.

Generally, a vague boundary exists between stationary and nonstationary point processes containing zonal stationary. The source of nonstationarity of these processes is the local regular behavior. These processes are further explained in Section \ref{Preliminaries}. 
Based on a point pattern, we want to decide between the zonal stationarity and the simple stationarity assumption. 
One may 
suggest to employ the methods developed for regular observation of random fields. To this end, since the point pattern is an irregular observation 
of points, we need to partition the observation window, $W$, into a regular $d$-dimensional grid where the $i$th edge of $W$ is 
divided into $n_{i}$, $i=1,\cdots,d$, equidistant parts. Let $N_{\mathbf{i}}$ represent the number of points in the sub-
cube in the $\mathbf{i}$th row, where $\mathbf{i}=(i_{1},\cdots, i_{d})$, and $i_{j}=1,\cdots,n_{j}$. We consider the 
$N_{\mathbf{i}}$ as a measured variable at the $\mathbf{i}$th grid and evaluate the stationarity of the point pattern by 
testing the stationarity of $\lbrace N_{\mathbf{i}}\rbrace_{\mathbf{i}}$ random field. This test is similar to the test for 
stationarity of random fields introduced by \cite{fuentes2005} which looses the information about the locations of 
points. 
In the second approach, we define the evolutionary spectra for the realization of a point process and then examine the local 
behavior of the discrete evolutionary spectra. The stationarity assumption is rejected if the spectral density function shows 
different behavior at least in two different regions of the window. We mention the property of different behaviors in different 
regions by \textit{second-order location dependency}. Our method is designed to detect this behavior against the second-
order stationarity.

The concept of evolutionary (i.e. time-dependent) spectra has been introduced by \cite{priestley, priestley1967}. 
Testing the stationarity of time series is a very famous consequences of this concept \cite{priestley1969}. The asymptotic properties of nonstationary time series with locally stationary behavior was investigated by \cite{dahlhaus} using the time-dependent spectra. The remaining references 
 somehow used the the evolutionary or local spectra for testing the stationarity of regular 
trajectories of random fields except \cite{bandyopadhyay} which proposed a test for spatial stationarity based on a transformation of irregularly spaced spatial data. A nonparametric and several parametric procedures for modeling the spatial dependence structure of a nonstationary spatial process observed on a $d$-dimensional lattice was proposed by \cite{fuentes2002a}. Priestley's idea was extended to test the stationarity and isotropy of a spatial 
process \cite{fuentes2005}. Fluctuations of the local variogram for testing the stationarity assumption of a random 
process was employed in \cite{corstanje}. A formal test for stationarity of spatial random fields on the bases of the asymptotic normality of certain covariance estimators at certain spatial lags was derived by \cite{Mikyoung}. A wavelet-based analysis of variance for a stationary random field on a regular lattice was used by \cite{Mondal} for testing the stationarity.

This paper is organized as follows: In the sequel, we study two motivating datasets. in the following section, we 
review some concepts of point processes and present two spatial spectral representations for a nonstationary 
spatial point process. Moreover, the nonparametric estimates of the spectral density of a nonstationary point process are proposed in this section. A formal test of stationarity is presented in Section \ref{test}. In Section \ref{simulation}, we examine the proposed testing approaches via a simulation study. We also study the effect of location in the competition between a specific genus of trees in \ref{realdata}.

\subsection{Data and motivations}
\begin{figure}
\begin{center}
\includegraphics[scale=0.5]{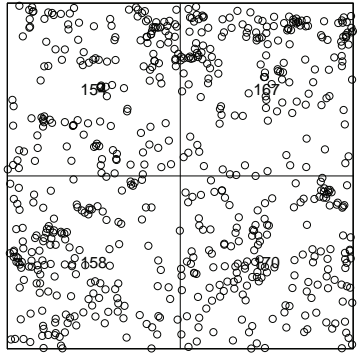}
\caption{The locations of $ 649 $ trees of the Bar Colorado forest in a rectangular window whose vertices are located 
at the geographic coordinates $(500, 0)$ and $(700,333/3333)$. 
The sampling window is rescaled into $ \left[ 0,70\right] ^{2}$.}\label{real data1}
\end{center}
\end{figure}

In this section, we propose two datasets as well as motivations for data analysis in order to further illustrate the mentioned 
problem throughout this study. The first dataset is devoted to the location of \textit{Euphorbiaceae} trees in a part of the 
Bar Colorado forest (Figure \ref{real data1}). Generally, one of the main and essential information for forest management and its optimal and 
sustainable utilization is to determine the frequencies of different types of trees, distribution pattern of trees, and their 
competition. For example, the distribution of trees seeds could have an impact on their aggregation while competition 
among different species for moisture, light, and nutrients acts vice versa. A positive autocorrelation or aggregation may 
result from regeneration near parents, whilst a negative autocorrelation results from their competition. Therefore, it seems 
that variations of the aggregation structure will be change the interaction structure of trees. Since the local periodogram is 
influenced by the aggregation and interactions of points, so we use this function for detecting the location(s) at which the 
structure of trees pattern is changed. 

\begin{figure}
\centering
    \begin{subfigure}[b]{0.3\textwidth}
        \includegraphics[scale=0.42]{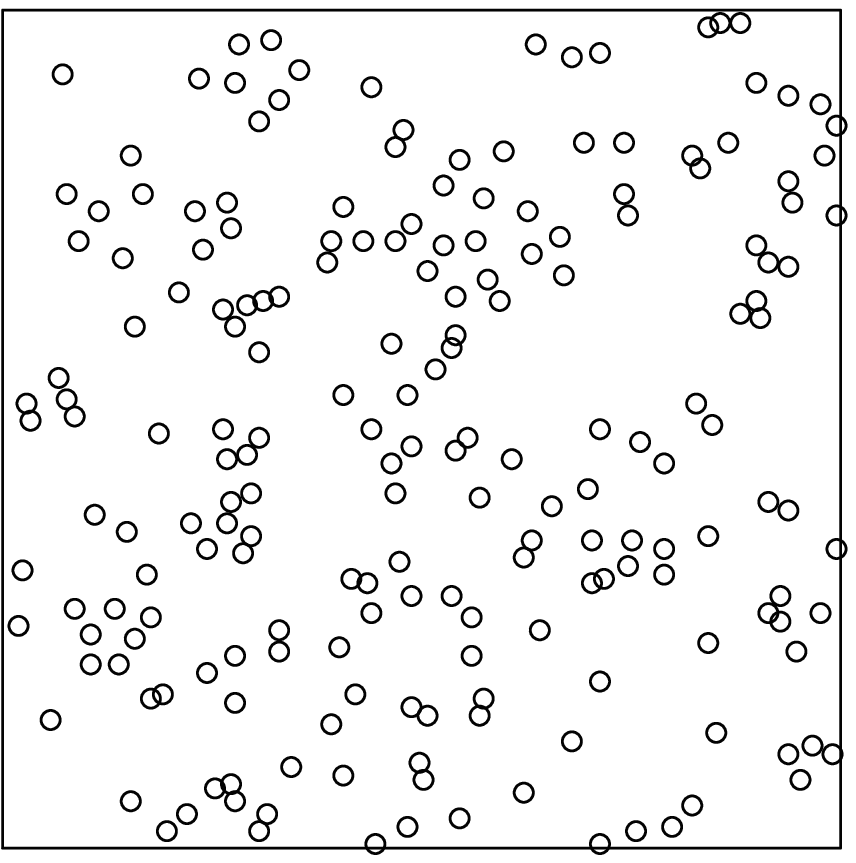}
        \caption{healthy tissue}\label{healthy}
    \end{subfigure}
    \begin{subfigure}[b]{0.3\textwidth}
        \includegraphics[scale=0.42]{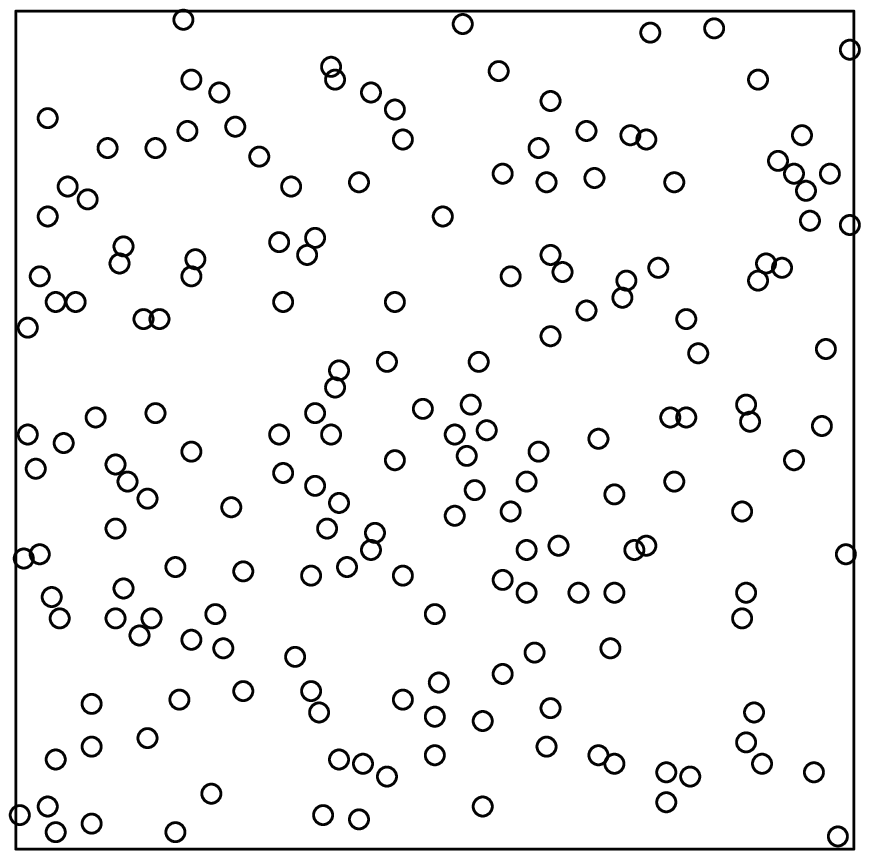}
        \caption{cancerous tissue}\label{cancerous}
    \end{subfigure}
\caption{The locations of $196$ capillaries on a section of a 
healthy prostate (\ref{healthy}) and the locations of $185$ capillaries at the same 
section from a cancerous prostate (\ref{cancerous}).}\label{real data2}
\end{figure}

The second data is the location of capillary profiles on a section of prostate tissue. According to the anatomy of the human body, blood circulation in the human body starts from heart, then flows through arteries, and finally receives to all parts of the body by capillaries. Capillaries are the smallest vessels of the body, supplying the oxygen to tissues and exchanging constituents between tissues and blood. Figure \ref{real data2} manifests the midpoints of the capillaries in sections of healthy and cancerous prostate tissues. The rescaled point patterns published by \cite{Hahn2012} were used to capture the coordinates of the capillaries. Since the quantity of nutrients and oxygen required by different parts of a given organ of the body are not same, so it is expected that the density of capillary network will be different in various parts of the body. In order to scrutinize this issue, we examine the behavior of evolutionary periodogram function at different parts of a tissue.
 
\section{Nonstationary spectral approaches}
\subsection{Preliminaries}\label{Preliminaries}
\begin{figure}[h]
\begin{center}
\includegraphics[scale=0.6]{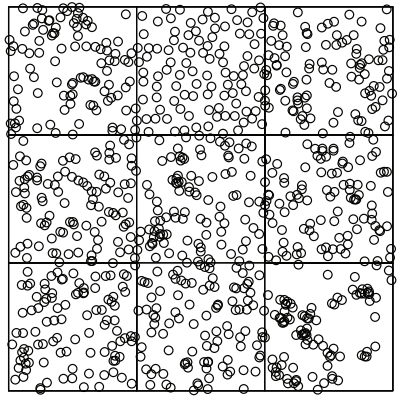}
\caption{A realization of the zonal stationary point process.}\label{fig1}
\end{center}
\end{figure}

A point process is called stationary if its distribution is invariant under translations and is called isotropic if its distribution is 
invariant under rotations about the origin in $\mathbb{R}^{d}$. For a stationary point process the intensity function 
is constant and the second-order characteristics depend only on the lag vector. 
Moreover, for an isotropic process such dependency is an exclusive function of the scalar length of lag vector 
regardless of the orientation. Moving to the frequency domain approaches, the spectral density function of 
a point process is the Fourier transform of the complete covariance density function. It is a function like 
$ \mathrm{f}_{XX}: \boldsymbol{\Omega} \longrightarrow \mathbb{C}$, where $\boldsymbol{\Omega}$ is the 
space of frequencies and is defined as (see \cite{mugglestone}) the Fourier transform of the unit-free complete covariance 
density function, $\kappa_{XX}$ \cite{Bartlett2}.
For a nonstationary spatial process, \cite{fuentes2005} generalized the concept of evolutionary spectra of time series 
introduced by \cite{priestley}. Following the same idea, we introduce two different evolutionary peridograms of 
nonstationary point processes. A new class of nonstationary point processes called `zonal' stationary processes is 
introduced. In each approach, an empirical spectral density function is defined for this class of nonstationary point process 
whose physical interpretation is similar to that of the spectrum of a stationary point process, but it is varying with location. 
Intuitively, a point process is called zonal stationary if it behaves in an approximately stationary way in a neighborhood of 
some point. In other words, a point process $ X $ is called zonal stationary on $W$ if there exist finite points $ 
\mathbf{z}\in W $ such that for some $ \rho>0 $, $ X_{b(\mathbf{z},\rho)} $ is stationary, where $ X_{B} $ denotes the 
restriction of $ X $ to $ B $ and $ b(\mathbf{z},\rho) $ denotes a closed ball centered at $ \mathbf{z} $ with a radius $ 
\rho $. Figure\ref{fig1} shows a realization of zonal stationary point process in the window $W=[0,70]^2$. We term $ 
z_{1},\ldots,z_{9} $ the middle points of $ \mathbf{S}_{1},\ldots,\mathbf{S}_{9} $ square sub-windows from left to right 
and down to up, respectively. In this figure, the point pattern in $ \mathbf{S}_{3}$ is a realization of the stationary 
Thomas 
process, in $ \mathbf{S}_{8}$ we have a realization of the Simple Sequential Inhibition (SSI) process and the remaining 
are realizations of stationary Poisson processes.


The regional behavior of a point pattern can be detected through evaluating its different features in the special regions. For this aim, closed balls with centers at different points of the observation window are typically considered in the spatial domain. Moreover, the local behavior of the point pattern is decided on the basis of measuring its different features at these balls. However, determining these regions and different features are crucial in the spatial domain. Fortunately, this issue can be easily solved in the frequency domain. This could be attributed to the fact that the frequency domain is based on $ \left[ -\pi,\pi\right] $ regardless of the form of the observation window. The Bartlett window can be used to determine different regions on which the features must be evaluated and accordingly the regional behavior of the point pattern can be detected by computing the local periodogram.

\begin{figure}[h]
\begin{center}
\includegraphics[scale=0.45]{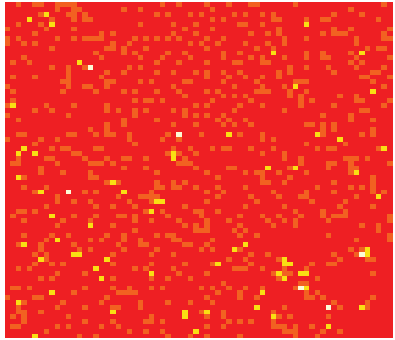}
\caption{The regular realization of the $N_{\mathbf{i}}$ random field corresponding to the mentioned point pattern in Figure \ref{fig1}.}\label{fig2}
\end{center}
\end{figure}

Suppose that we have a point process $X$ observed on a rectangular window $W=[\mathbf {0},\mathbf{l}]$, where $\mathbf{l}=(l_1,\ldots,l_d)$, that is the rectangular window with a vertex at the origin of the Cartesian coordinates and an opposite vertex on $\mathbf{l}$. Construct $\{N_{\mathbf{i}}\}_{\mathbf{i}}$ as before. 
We consider a regular $n_{1}\times \ldots \times n_{d}$ grid on the study area and set $n=\prod_{j=1}^{d}n_{j}$.  Let $N_{\mathbf{i}}$ be the number of points in the $\mathbf{i}$th row where $\mathbf{i}=(i_1,\ldots,i_d)$ and $i_j=1,\ldots,n_j$ for $j=1,\ldots,d$. We consider the $N_{\mathbf{i}}$ as a measured variable at the $\mathbf{i}$th quadrant and define the random field as $ V(\mathbf{s})=N_{\mathbf{i}} $ if $ \mathbf{s} $ belongs to the $\mathbf{i}$th row.  Figure \ref{fig2} shows the image of $N_{\mathbf{i}}$ random field obtained by considering a $ 70\times 70 $ regular grid on $y$.
Now, one may use the method of \cite{fuentes2005} to define the local periodogram of the obtained random field 
which eliminate the information about the locations. Our approach is the nonstationary periodogram. 

\subsection{Nonstationary periodogram of point processes}
Assume that we have a point pattern $x$ observed on a rectangular window $W=[\mathbf{0},\mathbf{l}]$ containing $ n_{X} $ points. Let $D=\lbrace \mathbf{u}_{j}=(u_{1j},\ldots,u_{dj}), j=1,\ldots, n_{X}\rbrace$, be the set of positions of the points. Suppose the approximate locations of points in a point pattern be represented by the intersections of a $ m_{1}\times \ldots \times m_{d} $ fine lattice superimposed on the study region. This lattice generates an irregular realization of a binary random field. We define the obtained realization, $\zeta$ say, from the random field $ \boldsymbol{\zeta} $ as 
\begin{eqnarray*}
\zeta(\mathbf{s})=\left\{
\begin{array}{rl}
1 ~~& \text{if }~ \mathbf{s}\in D\\
0 ~~& \text{if } ~\mathbf{s} \notin D.
\end{array} \right.
\end{eqnarray*}
$ \boldsymbol{\zeta}(\mathbf{s}) $ is a simple random field and only the positions at which events occur are of interest. 
There is an equivalence between the point spectra of $ X $ and the spectra of the random field $\boldsymbol{\zeta} $ 
\cite{renshaw1}. If $ X $ is a stationary point process, then $\boldsymbol{\zeta} $ will be a stationary random field. 
Consequently, any evidence of nonstationarity of $\boldsymbol{\zeta}$ could be a reason for rejection of the stationarity 
hypothesis of $X$. 
Suppose that $ \boldsymbol{\zeta} $ be zonal stationary, then its spectral density function, denoted by $\mathrm{f}_{z}(\boldsymbol{\omega})$, is the Fourier transform of the locally auto covariance function of $ \boldsymbol{\zeta}  $ and 
$ z $ is the location around that $ \boldsymbol{\zeta} $ behaves in a stationary manner (see \cite{fuentes2005}). 
If for a given $ z$ the function $ \mathrm{f}_{z}
(\boldsymbol{\omega}) $ is not sensitive to $ \boldsymbol{\omega} $, then we conclude that in the neighborhood of the 
given $ z $, $ \mathrm{f}_{z}(\boldsymbol{\omega}) $ belongs to a completely spatial random point pattern. On the other 
hand, if for a given frequency $ \boldsymbol{\omega} $, $ \mathrm{f}_{z}(\boldsymbol{\omega}) $ is not sensitive to $ z 
$, then we conclude that for the given $ \boldsymbol{\omega} $, $ \mathrm{f}_{z}(\boldsymbol{\omega}) $ dose not 
behave locally. Practically, $ \mathrm{f}_{z}(\boldsymbol{\omega}) $  is unknown for every $ z $ and for every 
$\boldsymbol{\omega} $ and we restrict the problem of estimation to specific Fourier frequencies and locations. 
Therefore, our approach is a parametric statistical inference.

Consider a location varying filter to assign greater weights to the neighboring values of $ z $. The spatial local periodogram at a location $z$ and frequency $\boldsymbol{\omega}$ is $\vert \mathcal{J}_{z}(\boldsymbol{\omega})\vert^{2}$, where
\begin{eqnarray}\label{second estimator}
\mathcal{J}_{z}(\boldsymbol{\omega})=&\frac{1}{\sqrt{\prod_{j=1}^{d}l_{j}}}\sum_{k=1}^{m}g(z-\mathbf{s}_{k})\zeta(\mathbf{s}_{k})\exp\lbrace -i\mathbf{s}_{k}^{T}\boldsymbol{\omega} \rbrace \nonumber \\ =&
\frac{1}{\sqrt{\prod_{j=1}^{d}l_{j}}}\sum_{j=1}^{n_{X}}g(z-\mathbf{u}_{j})\exp\lbrace -i\mathbf{u}_{j}^{T}\boldsymbol{\omega} \rbrace \nonumber \\
=&A_{z}(\boldsymbol{\omega})+iB_{z}(\boldsymbol{\omega}),
\end{eqnarray}
where $g:\mathbb{R}^d\rightarrow\mathbb{R}$ is the filter function with all the characteristics mentioned by 
\cite{fuentes2005} with finite width $B_g$ defined as 
$B_{g}=\int_{\mathbb{R}^{d}} \vert \mathbf{u} \vert \vert g(\mathbf{u}) \vert \mathrm{d}\mathbf{u}$, 
and $m=m_1\times\ldots\times m_d$. The resulting periodogram is not smooth enough to be used as an estimation 
of the local spectra and thus $|\mathcal{J}_z|^2$ is smoothed using the $\mathbb{L}^2$ kernel family, 
$\{W_{\rho}\}$, where for each $\rho$ there exists a constant $C$ such that 
\begin{eqnarray}\label{C}
\lim_{\rho\rightarrow \infty} \rho^{d}\int_{\mathbb{R}^d} \vert w_{\rho}(\mathbf{\lambda})\vert^{2} \mathrm{d}\mathbf{\lambda}=C,
\end{eqnarray}
where $w_{\rho}$ is the Fourier transform of $W_{\rho}$. Thus, the final estimator is 
\begin{eqnarray}\label{1}
\widehat{f}_{z}(\boldsymbol{\omega})=I_{z}(\boldsymbol{\omega})= \int_{\mathbb{R}^d} W_{\rho}(z-\mathbf{u}) \vert \mathcal{J}_{z}(\boldsymbol{\omega})\vert^{2} \mathrm{d}\mathbf{u}.
\end{eqnarray}
Using the very similar arguments employed by \cite{fuentes2005}, the covariance between the spatial periodogram values $I_{z_{1}}(\boldsymbol{\omega})$ and $I_{z_{2}}(\boldsymbol{\omega}^{'})$ will be asymptotically zero if either $\Vert \boldsymbol{\omega}\pm \boldsymbol{\omega}^{'} \Vert \gg$
bandwidth of $ \vert \Gamma(\boldsymbol{\theta})\vert^{2}$ where $\Gamma$ is the Fourier transform 
of $g$ and or $\Vert z_{1}\pm z_{2}\Vert \gg$ bandwidth of the function $ W_{\rho}(\mathbf{u})$ \cite{priestley1966}. For fixed $z$ and $\boldsymbol{\omega}$, one may conclude the normality of $A_{z}(\boldsymbol{\omega}) $ and $B_{z}(\boldsymbol{\omega})$ similar to the \cite{mugglestone}.

\begin{figure}[h]
\begin{center}
\includegraphics[scale=0.2]{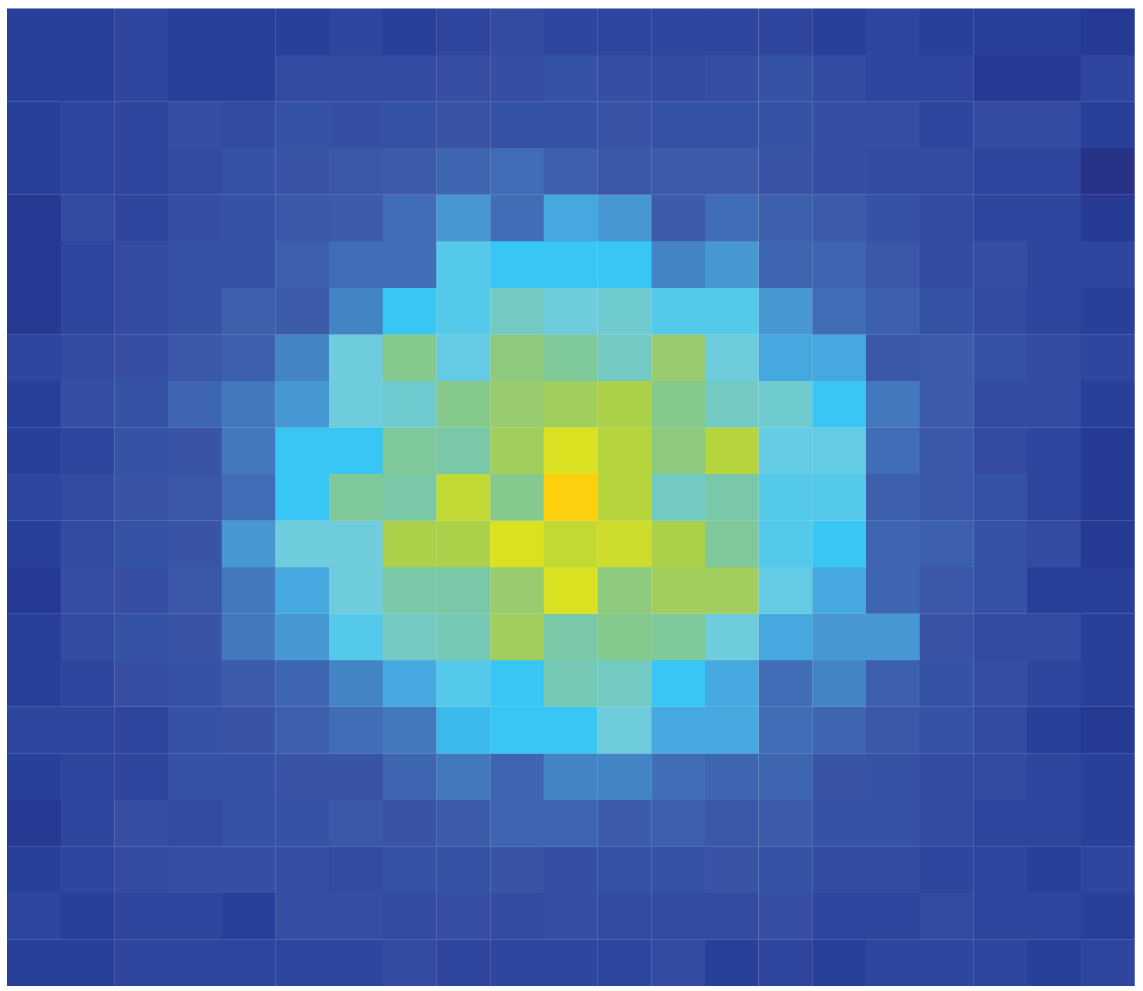}
\includegraphics[scale=0.2]{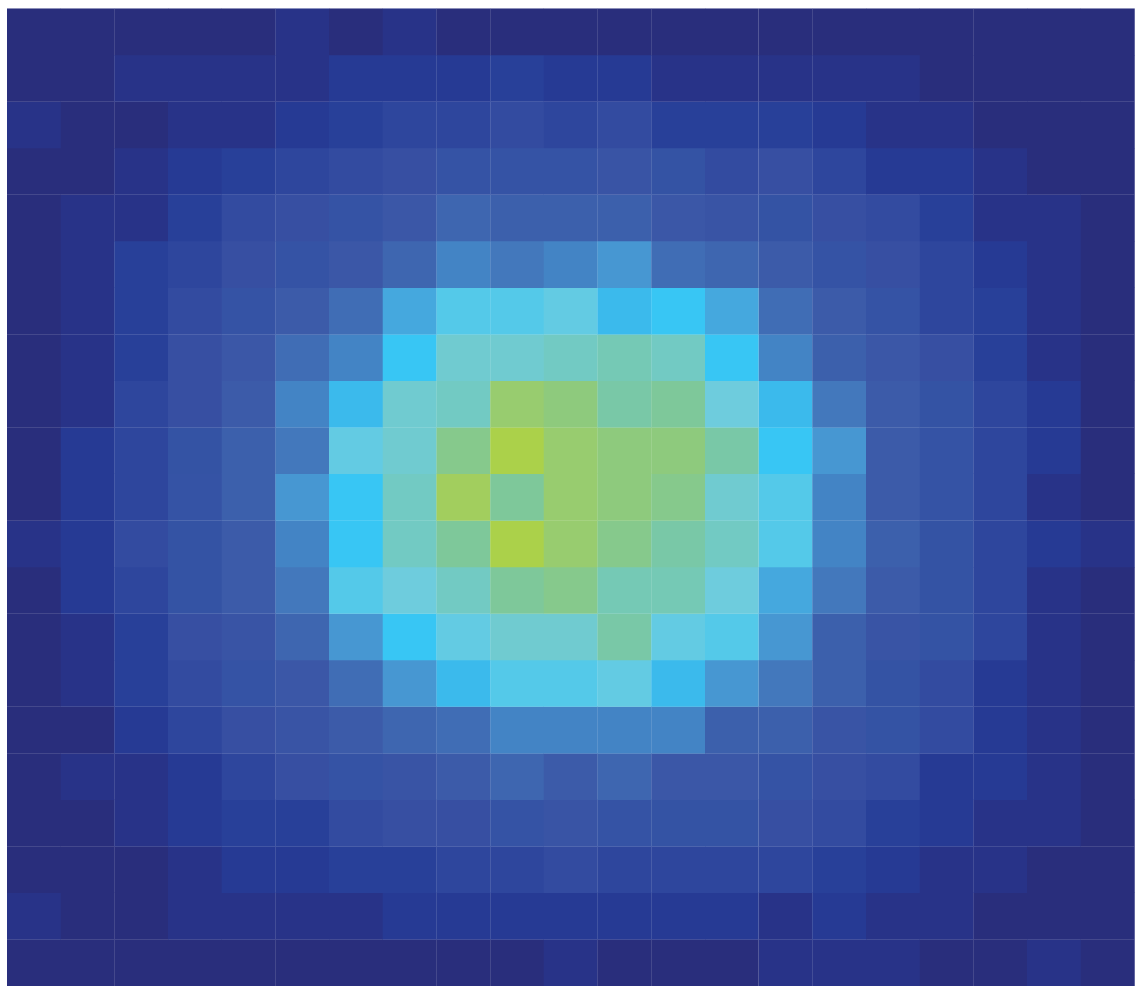}
\includegraphics[scale=0.2]{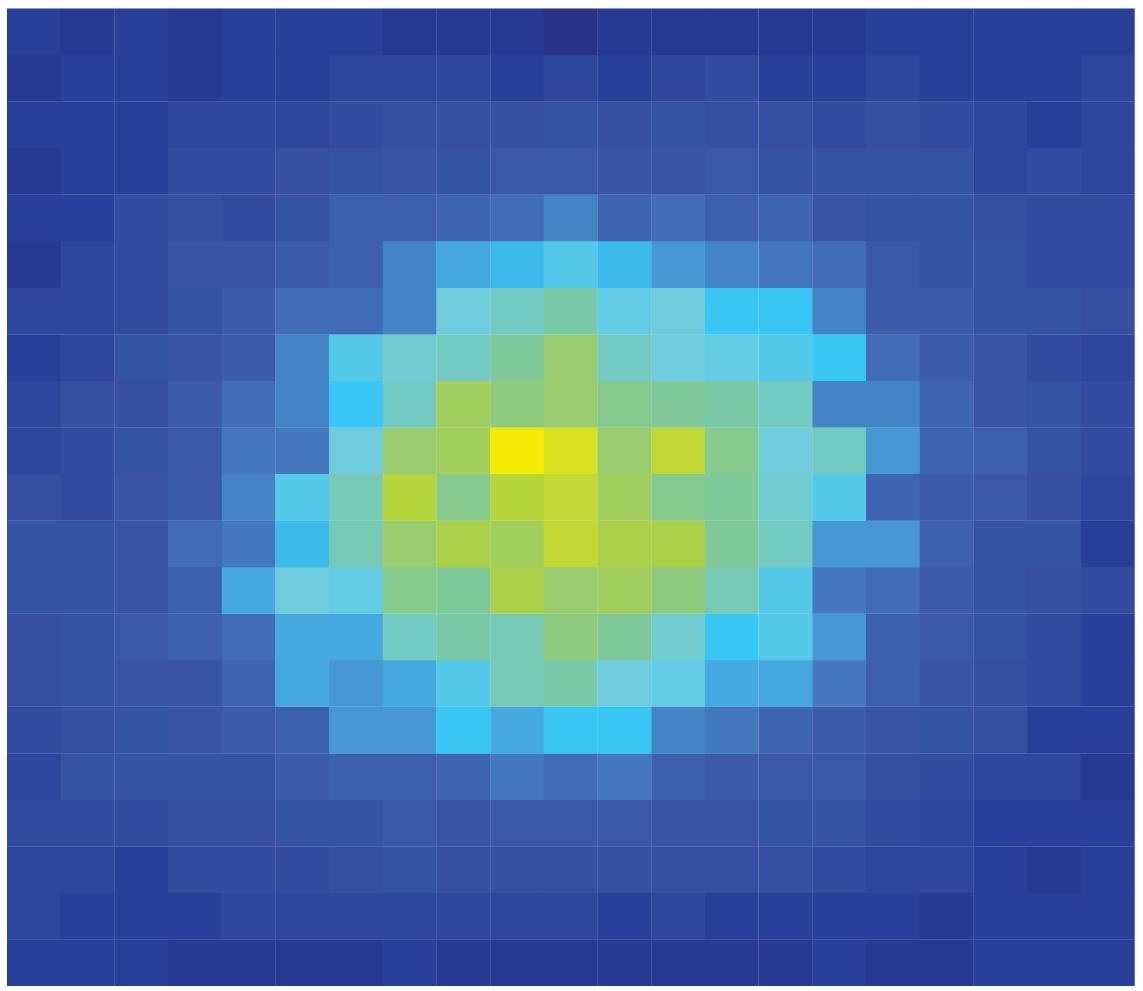}\\
\includegraphics[scale=0.2]{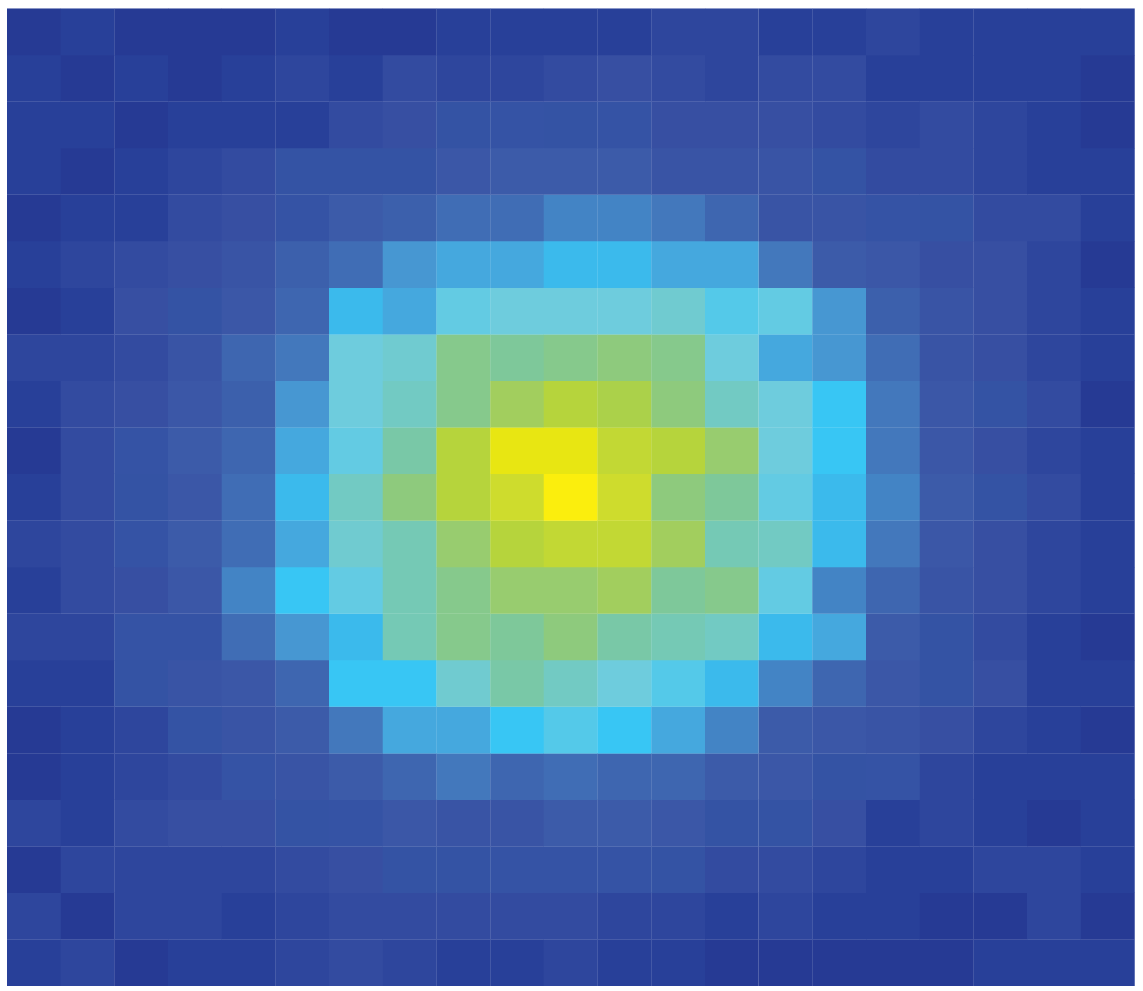}
\includegraphics[scale=0.2]{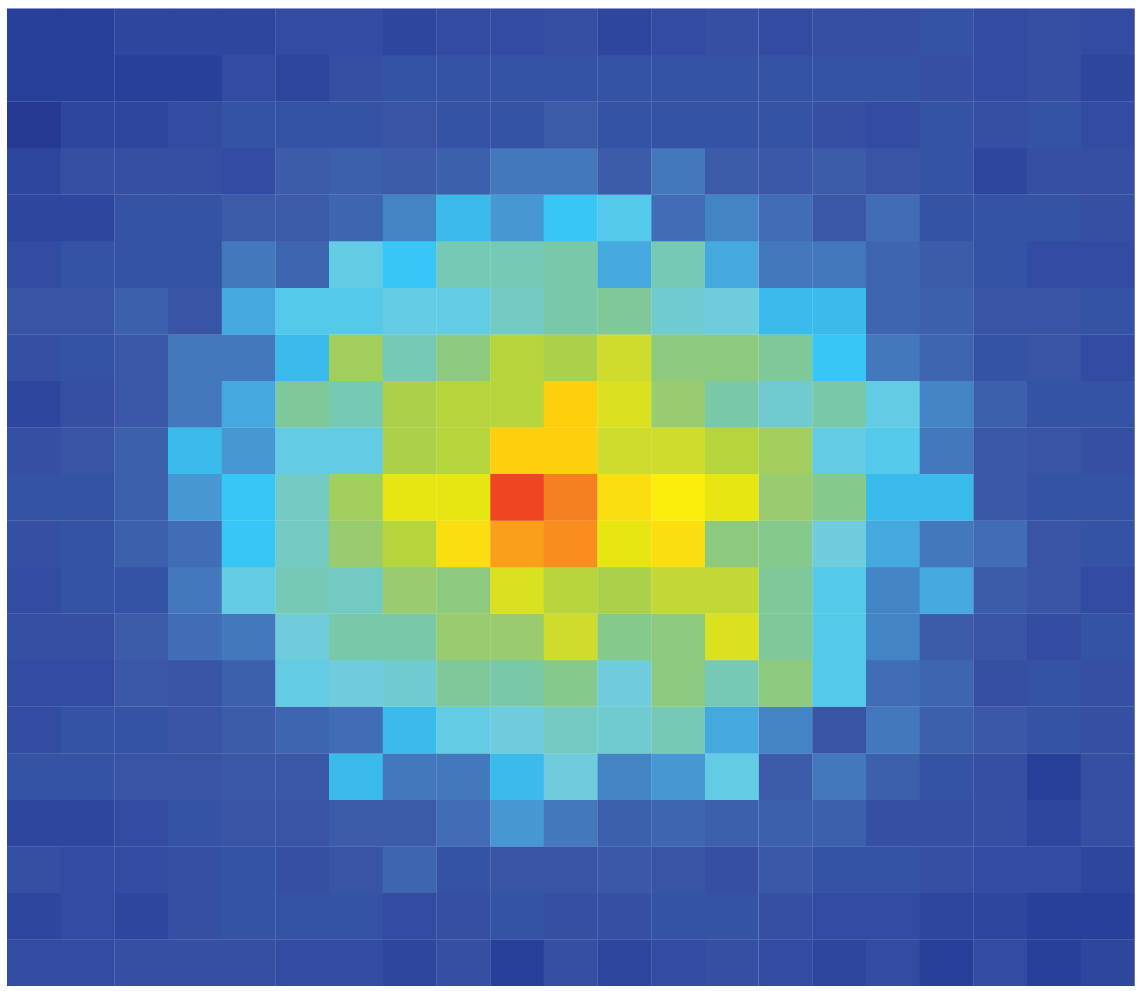}
\includegraphics[scale=0.2]{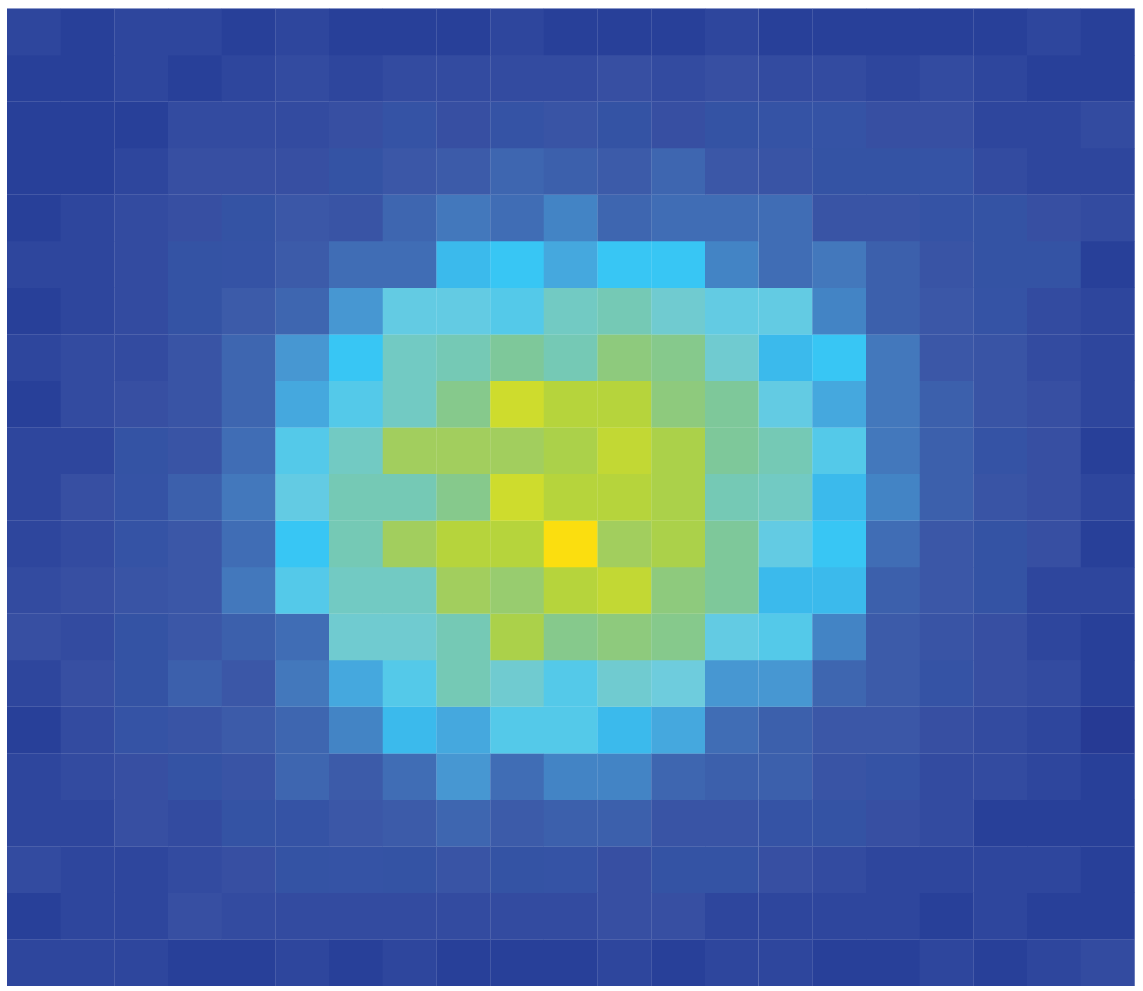}\\
\includegraphics[scale=0.2]{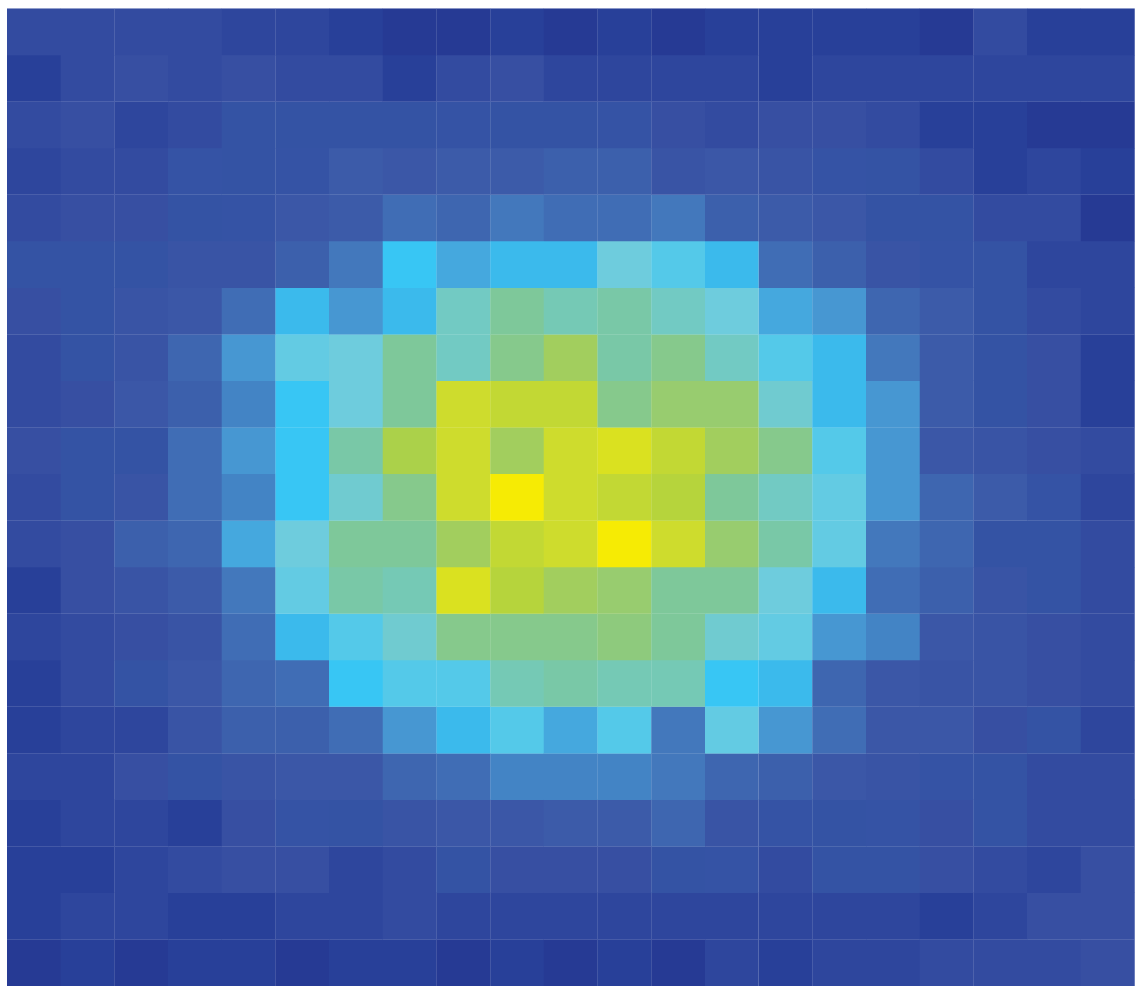}
\includegraphics[scale=0.2]{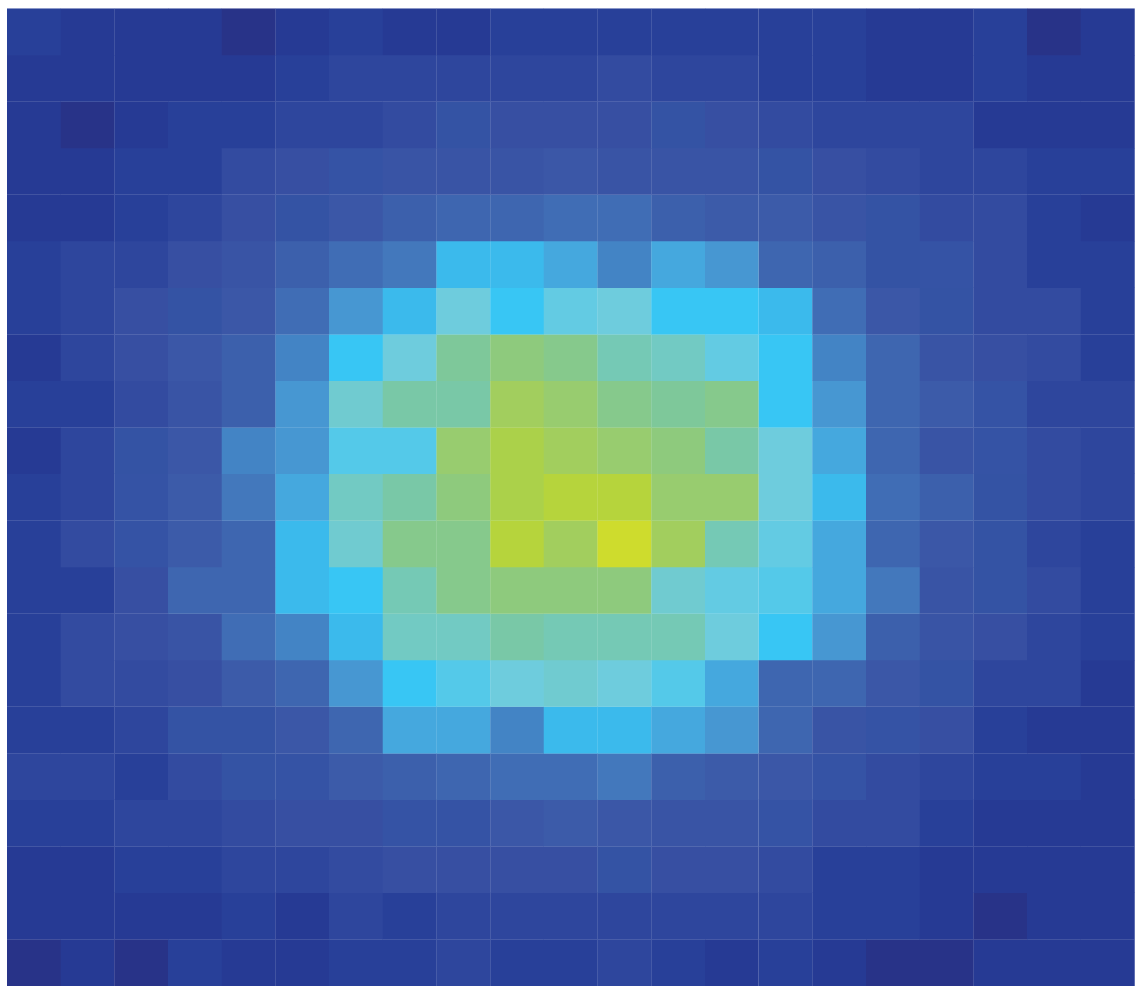}
\includegraphics[scale=0.2]{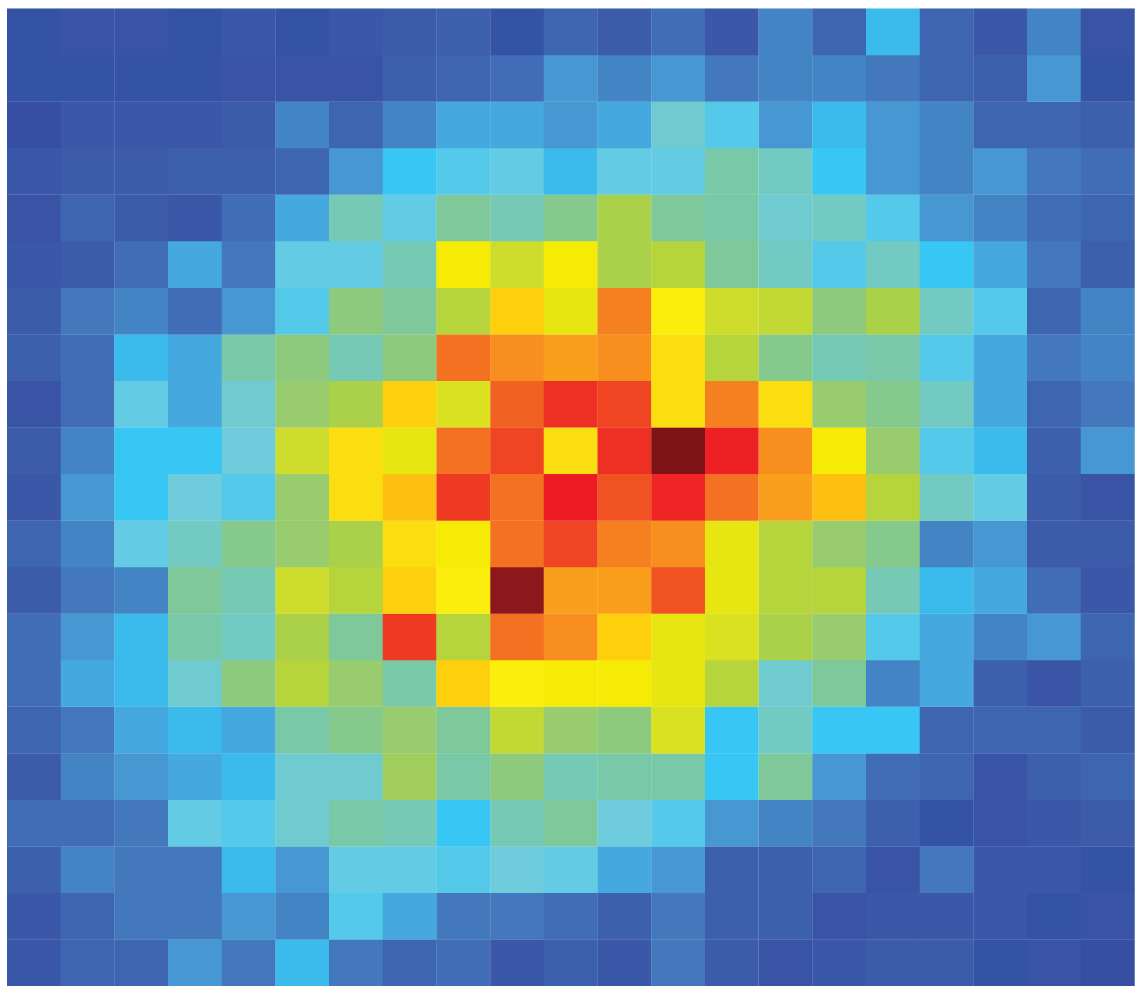}\\
\includegraphics[scale=0.45]{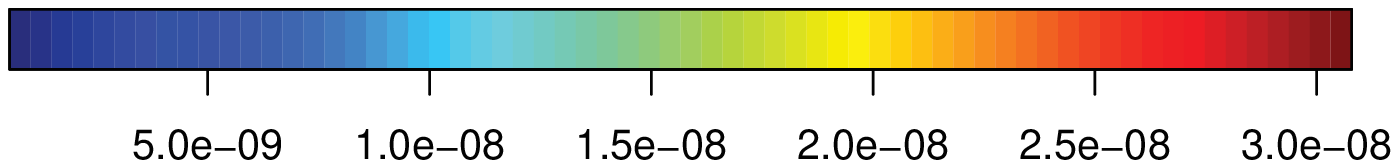}
\caption{The spatial local periodogram of the point pattern of Figure \ref{fig1} at the locations $z_{1},\ldots,z_{9}$ and Fourier frequencies }\label{fig3}
\end{center}
\end{figure}

Figure \ref{fig3} shows the spatial local periodogram of the mentioned $ y $ in Figure \ref{fig2} at the locations $ z_{1},\ldots,z_{9} $ which are enough wide apart and at the Fourier frequencies. 
The number of observed points of $ y $ in the considered sub-windows are almost similar. Obviously, as shown in Figure \ref{fig3}, the behavior of the local periodogram function of $ y $ varies at different locations. The shape of the local periodograms in sub-windows with stationary Poisson patterns (around $z_k$, $k\in K=\{1,2,4,5,6,7,9\},$) are broadly flat, reflecting the absence of the interaction structure in the observed patterns. In the sub-window with the stationary Thomas pattern, for small values of $\| \boldsymbol{\omega}\|$ the values of $I_{z_{3}}(\boldsymbol{\omega})$ are larger in comparison with $I_{z_j}(\boldsymbol{\omega})$ when $j\in K$. In contrast, the sub-window with the Simple Sequential Inhibition pattern, the values of $I_{z_{8}}(\boldsymbol{\omega})$ are smaller than $I_{z_j}(\boldsymbol{\omega})$, $j\in K$, for small $\|\boldsymbol{\omega}\|$.

There is a fundamental difference between this location dependent point spectra and the lattice-based local spectra considered by \cite{fuentes2005}. A lattice-based spectrum describes the spatial structure of a measured variable (for example, tree heights) at fixed equally spaced locations in an $n_{1}\times n_{2}$ grid. However, in a local point spectrum, we consider the spatial structure of the location (for example, locations of trees) of points instead of a measured characteristic. Moreover, the first estimator ignores the information about the precise locations of points compared with the second one.

%
%
%
Concerning the Bartlett's window, $ g(\mathbf{u})$ is considered as a multiplicative filter in the simulation study of Section \ref{simulation}; i.e. $g(\mathbf{u})=\prod _{j=1}^{d}g_{1}(u_{j})$, where 
$g_{1}(u)=(4\pi h)^{-1/2}\mathbf{1}_{|u|\leq h}$. Thus the Fourier transform of $g$ becomes 
$\Gamma(\boldsymbol{\omega})=\prod_{j=1}^{d}(\sqrt{h\pi} \omega_{j})^{-1}\sin(h\omega_{j})$. 
Further, we consider $W_{\rho}(\mathbf{u})$ to be of the form $W_{\rho}(\mathbf{u})=\prod _{j=1}^{d}W_{1,\rho}(u_{j})$, where $W_{1,\rho}(u)=1/\rho\mathbf{1}_{|u|\leq\rho/2}$, 
corresponding to the Daniell window
which implies the accuracy of formula \eqref{C} with $C=(2\pi)^{d}$.

\section{Testing the stationarity}\label{test}
In this section, we represent a formal test of stationarity of a point process based on the results discussed in the previous 
section.The analysis of the geometric structure of a point pattern is strongly related to the aggregation of points and 
interaction between points. When the aggregation of points is the same at two different sub-windows it is not easy to 
visually discriminate the interaction effect in these sub-windows. In this situation, the higher moments of the point pattern 
such as the second order intensity function must be compared together in different areas. 
If the restricted point patterns  to 
disjoint sub-windows are assumed to be independent, their second order properties can be compared according to 
\cite{saadat}. Without such independence, the \textit{two-factor analysis of variance} model and the assumptions 
considered in the following are valid.
We thus wright (see also \cite{jenkins1961,priestley1969}) 
\begin{eqnarray}\label{Y}
Y(z,\boldsymbol{\omega})=\ln I_{z}(\boldsymbol{\omega})=\ln f_{z}(\boldsymbol{\omega})+\varepsilon(z,\boldsymbol{\omega}).
\end{eqnarray}
\cite{priestley1969} showed that for $\boldsymbol{\omega}\in \prod_{j=1}^{d} \left[ -\pi/\Delta_{j},\pi/\Delta_{j}\right]$, the asymptotic mean and variance of $\varepsilon$ are $\mathbb{E}(\varepsilon(z,\boldsymbol{\omega}))=0$ and 
\begin{eqnarray}\label{sigma}
\sigma^{2}=\left( C/\rho^{d} \right) \int_{\mathbb{R}^d} \vert \Gamma(\boldsymbol{\theta})\vert^{4} \mathrm{d}\boldsymbol{\theta},
\end{eqnarray}
which is independent of $z$ and $\boldsymbol{\omega}$.

We select a set of locations $z_{1}, z_{2},\ldots, z_{m}$ and a set of frequencies $\boldsymbol{\omega}_{1}, \boldsymbol{\omega}_{2},\ldots, \boldsymbol{\omega}_{n}$ in such a way that an acceptable sample is gathered in location and frequency spaces and we evaluate $I_{z}(\boldsymbol{\omega})$ over this fine sample. The set of locations and frequencies must also be wide apart enough in order to $\varepsilon(z_{i},\boldsymbol{\omega}_{j})$ and consequently $I_{z_{i}}(\boldsymbol{\omega}_{j})$ be uncorrelated. 
If we write $Y_{ij}=Y(z_{i},\boldsymbol{\omega}_{j})$, $f_{ij}=\ln f_{z_{i}}(\boldsymbol{\omega}_{j})$ and $\varepsilon_{ij}=\varepsilon(z_{i},\boldsymbol{\omega}_{j})$, then using the discussion on the normality of $A_z$ and $B_z$, the noise terms, $\varepsilon_{ij}$, are considered to be normally distributed and $Y_{ij}$ are generated according to the model 
\begin{eqnarray*}
Y_{ij}= f_{ij}+\varepsilon_{ij}.
\end{eqnarray*}
The parameter $f_{ij}$ represents the treatment effect for location effect at level $z_i$ and the frequency effect at level $\boldsymbol{\omega}_{j}$. By considering the normality of the main effects, we can rewrite the model as
\begin{eqnarray}\label{H1}
H_{1}:Y_{ij}=\mu+\alpha_{i}+\beta_{j}+\gamma_{ij}+\varepsilon_{ij}, i=1,\ldots,m\quad \text{and}\quad j=1,\ldots,n.
\end{eqnarray}
In this model, the parameters $\alpha_{i}$ and $\beta_{j}$ represent the main effects of the location and frequency 
factors, respectively, and $\gamma_{ij}$ represents the interaction between these two factors. 
As previously 
mentioned, the spectral density function of a stationary point process does not oscillate locally and it only varies by 
change of the frequencies. 
Therefore, the stationarity of a point process can be tested by using the analysis of variance 
methods and testing the model 
\begin{eqnarray*}
H_{0}:Y_{ij}=\mu+\beta_{j}+\varepsilon_{ij},~ i=1,\ldots,m\quad \text{and}\quad j=1,\ldots,n ,
\end{eqnarray*}
or equivalently $H_0: \alpha_{i}=0,~ i=1,\ldots,m,$ versus (\ref{H1}). The rejection of $H_{0}$ represents that at least 
one of the parameters $\alpha_{i}$ is not zero which means that there exists at least one $z_{i}\in W $ in such a way that 
$f_{z_i}(\boldsymbol{\omega})\neq f_{z_j}(\boldsymbol{\omega})$ for all $j\neq i$. A post-hoc test is used to find the 
different location(s) and thus we may use this as a clustering method in zonal stationary point processes. Since the value of 
$\sigma^{2}=\mathbb{V}ar(\varepsilon_{ij})$ is known, the presence of the interaction factor, $\gamma_{ij}$, can be 
tested only using one realization of the point process. If the effect of interaction factor is not significant, then the point 
process is uniformly modulated and $\ln f_{z}(\boldsymbol{\omega})$ will be additive in terms of location and frequency. 
We can examine whether the nonstationarity of the point pattern is restricted only to some frequencies, by selecting those 
frequencies and testing for stationarity over them. If $X$ is an isotropic point process, then 
$f_{z}(\boldsymbol{\omega})$  depends on its vector argument $\boldsymbol{\omega}$ only through its scalar 
length $\Vert\boldsymbol{\omega}\Vert$, regardless of the orientation of $\boldsymbol{\omega}$. Then, we can test 
for isotropy by selecting a set of frequencies with the same norms, say 
$\lbrace \boldsymbol{\omega}_{j_{1}},\boldsymbol{\omega}_{j_{2}}\rbrace$ where 
$\boldsymbol{\omega}_{j_{1}}\neq\boldsymbol{\omega}_{j_{2}}$ but 
$\Vert \boldsymbol{\omega}_{j_{1}}\Vert=\Vert\boldsymbol{\omega}_{j_{2}}\Vert$, and test whether the 
frequency effect is significant. Since $\sigma^{2}$ is known, all of  these comparisons are based on a $\chi^{2}$ rather 
than $F$-test.

\section{Simulation study}\label{simulation}
In this section, we evaluate the performance of the proposed test for detecting the nonstationarity of a point process. As mentioned in the previous section, using a logarithmic transformation, the mechanism of the test are almost identical to those of a two-factor analysis of variance procedure. 

In testing for stationarity, firstly we consider the interaction sum of squares. If the interaction is not significant, we conclude 
that the point process is a uniformly modulated process and continue to test for stationarity by evaluating the `between 
spatial locations' sum of squares. If the interaction sum of squares turns out to be significant, we conclude that the point 
pattern is nonuniformly modulated and nonstationary. In this situation, we can survey whether the nonstationarity of the 
point process is restricted only at some frequencies. Significance of the `between spatial locations' sum of squares means 
that the point pattern is nonstationary. 

We set the nominal level of test at $0.05$ and assume that all of the point processes are observed on a rectangular window 
$W=\left[ 0,70\right] ^{2}$. Using \eqref{second estimator}, for all cases, we estimate $f_{z}(\boldsymbol{\omega})$ in 
which $g(\mathbf{u})$ and $W_{\rho}(\mathbf{u})$ employ $h=3$ and 
$\rho=20$, respectively. The bandwidth of $\vert \Gamma(\omega_{1})\vert^{2}$ is approximately $\pi/h=\pi/3$. The window 
$W_{\rho}(\mathbf{u})$ has a bandwidth of $\rho=20$. Thus, the space between $z_{i}$ and 
$\boldsymbol{\omega}_{j}$ points must be at least $\rho=20$ and $\pi/3$ respectively, in order to obtain approximately 
uncorrelated estimates. The points $z_{i}$ are chosen as $z_{i}=(z_{i_{1}},z_{i_{2}})=(70i_{1}/6,70i_{2}/6)$ with 
$i_{1}=1 (2) 5$ and $i_{2}=1 (2) 5$, corresponding to a uniform spacing of $70/3$ (just exceeding $\rho=20$). The 
frequencies are chosen as $ \boldsymbol{\omega}_{1}=( \pi/20, \pi/20)$, 
$  \boldsymbol{\omega}_{2}=( 8\pi/20, \pi/20) $, $  \boldsymbol{\omega}_{3}=( 15\pi/20, \pi/20) $, 
$  \boldsymbol{\omega}_{4}=( \pi/20, 8\pi/20) $, 
$  \boldsymbol{\omega}_{5}=( 8\pi/20, 8\pi/20) $, $  \boldsymbol{\omega}_{6}=( 15\pi/20, 8\pi/20) $, 
$  \boldsymbol{\omega}_{7}=( \pi/20, 15\pi/20) $, $  \boldsymbol{\omega}_{8}=( 8\pi/20, 15\pi/20) $ and 
$  \boldsymbol{\omega}_{9}=( 15\pi/20, 15\pi/20) $, with respect to the regular spacing of $7\pi/20$ (just exceeding 
$\pi/3$). The value of $\sigma^{2}$ is calculated using equation \eqref{sigma} as 
$\sigma^{2}=16h^{2}/(9\rho^{2})=0.04$. For the considered set of locations and frequencies, the degree of freedom 
for `between spatial locations', `between frequencies' and `residual+ interaction' effects are 
$ df_{L}=8 $, $ df_{F}=8 $ and $df_{IEr}=64 $, respectively. We denote `between spatial locations', 
`between frequencies' and `residual+ interaction' sum of squares with $ SSL $, $ SSF $ and $ SSIEr $, respectively. In 
the analysis of variance table, we first evaluate the significance of the interaction effect. If 
$ SSIEr/ \sigma^{2} > \chi^{2}_{64}(0.05) $, we conclude that the point pattern is 
nonstationary, and nonuniformly modulated. If the interaction is not significant, we conclude that the point pattern is a 
uniformly modulated process, and proceed to test for stationarity versus the zonal stationarity by comparing $ SSL/ 
\sigma^{2}$ with $ \chi^{2}_{8}(0.05) $. Significance of the location effect suggests that the point pattern is 
nonstationary. Similarly, to evaluate between frequencies effect, $ SSF/ \sigma^{2}$ is compared with $ \chi^{2}_{8}
(0.05) $ and the significance of test confirms that the spectra is nonuniform.

In the following study, we simulate $ 100 $ realizations from a stationary and a nonstationary point process to study the 
empirical size and power of the proposed test. Table \ref{tab1} shows the ratios of rejections of stationarity assumption. 
The 
ratios of rejections for stationary processes represent the empirical size and the ratios of rejections for nonstationary 
processes represent the empirical power of the test. We consider realizations of Thomas process because of its clustered 
behavior which makes it similar to a zonal stationary process. To simulate a Thomas process of the parameter $ (\delta,\tau,
\mu) $, the `parent' point process is generated at first according to a stationary Poisson point process of intensity $\delta$ 
in the study region. Then, each parent point is replaced independently by `offspring' points which the number of offspring 
points of each parent is generated according to a Poisson distribution with mean $\mu$. Finally, the position of each 
offspring relative to its parent location is determined by a bivariate normal distribution centered at the location of parent 
with covariance matrix $diag(\tau^{2} ,\tau^{2}) $. The process is denoted by $T(\delta,\tau,\mu)$ and is chosen because of its clustered 
behavior which makes it similar to a zonal stationary process. The number of observed points is an increasing function of $ \delta $ 
in Thomas process. We also consider the realizations of stationary Poisson processes of intensity $\lambda$, denoted by $P(\lambda)$.  The results of Table \ref{tab1} show 
that the empirical size of test decreases for the larger values of $ \delta $ and approaches the nominal level. The great 
values of empirical size are due to the use of asymptotic normal distribution for $Y_{ij}$. In fact, this asymptotic behavior 
is 
valid for large enough number of points at each grid. For different realizations of stationary Poisson processes, the 
empirical 
size is zero. For the point pattern $y$ shown in Figure \ref{fig1}, the point pattern in $ \mathbf{S}_{3}$ is a realization of 
the stationary Thomas process with parameters $\delta=0.046$, $\tau=1$ and $\mu=4$, and the point pattern in $ 
\mathbf{S}_{8}$ is a realization of the SSI process with Inhibition distance $r=1.5$. 

To simulate a SSI process with Inhibition distance $r$, points are added one-by-one. Each new point is generated according 
to uniform distribution in the window and is independent of previous points. If the new point lies closer than $r$ from an 
existing point, then it is rejected and another random point is generated. The empirical power of test is close to one for the 
point pattern $y$.
The empirical power of test is equal to one in all the nonstationary cases. 

Since the value of the periodogram at higher frequencies can be taken as the contribution of random errors only, so we ignore $ \boldsymbol{\omega}_{9}=( 15\pi/20, 15\pi/20) $ (the frequency with the highest norm) and perform the test by considering $ \boldsymbol{\omega}_{1},\ldots,\boldsymbol{\omega}_{8} $. Therefore, we have $ df_{F}=7 $ and $ df_{IEr}=56 $. The results of Table \ref{tab1} show that the empirical size of test decreases when $ \boldsymbol{\omega}_{9} $ is removed from the set of considered frequencies.

\begin{table}
\caption{\small The rejections ratios of stationarity in $100$ times replications of testing procedure with the realization of Thomas and Poisson point processes.}
\label{tab1}
\vspace{.25cm}
\centering
\footnotesize
\scalebox{0.8}{%
\begin{tabular}{lcc}
\hline
Model & using 9 frequencies& without $ \boldsymbol{\omega}_{9} $ \\
\hline
$T$(2,1,6)			&0.10	&0.04	\\	
$T$(2,1,8)			&0.13	&0.08	\\	                   
$T$(5,0.25,4)		&0.12	&0.02	\\	  
$T$(5,0.25,6)		&0.06	&0.05	\\	         
$T$(5,0.25,8)		&0.06	&0.03	\\	                           
$T$(5,0.5,4)		&0.02	&0.01	\\	         
$T$(5,0.5,6)		&0.00	&0.01	\\	    
$T$(5,0.5,8)		&0.05	&0.01	\\	                 
$T$(5,1,4)			&0.03	&0.02	\\ 	 	
$T$(5,1,6)			&0.00	&0.00	\\
$T$(5,1,8)			&0.00	&0.00\\
$T$(7,0.25,4)		&0.06	&0.02\\ 
$T$(7,0.25,6)		&0.02	&0.00\\ 
$T$(7,0.25,8)		&0.04	&0.02\\ 
$T$(7,0.5,4)		&0.01	&0.00\\ 
$T$(7,0.5,6)		&0.02	&0.01\\ 
$T$(7,0.5,8)		&0.00	&0.00\\ 
$T$(7,1,4)			&0.00	&0.00\\ 
$T$(7,1,6)			&0.00	&0.00\\ 
$T$(7,1,8)			&0.00	&0.00\\ 
$P$(20)				&0.00	&0.00\\
$P$(30)				&0.00	&0.00\\
$P$(50)				&0.00	&0.00\\ 
$P$($0.1\exp(0.2\sin(4\pi x)+0.1y)$)	&1.00 &1.00 \\ 
$T$(3, 0.2,$\exp(3\sin(4\pi xy)+0.01x)$)		&1.00&1.00\\
$y$&0.98&1.00\\ 
\hline
\end{tabular}
}
\end{table}

\subsection{Real data}\label{realdata}
\subsubsection{Trees data}

\begin{figure}[h]
\centering
    \begin{subfigure}{0.4\textwidth}
        \includegraphics[scale=0.3]{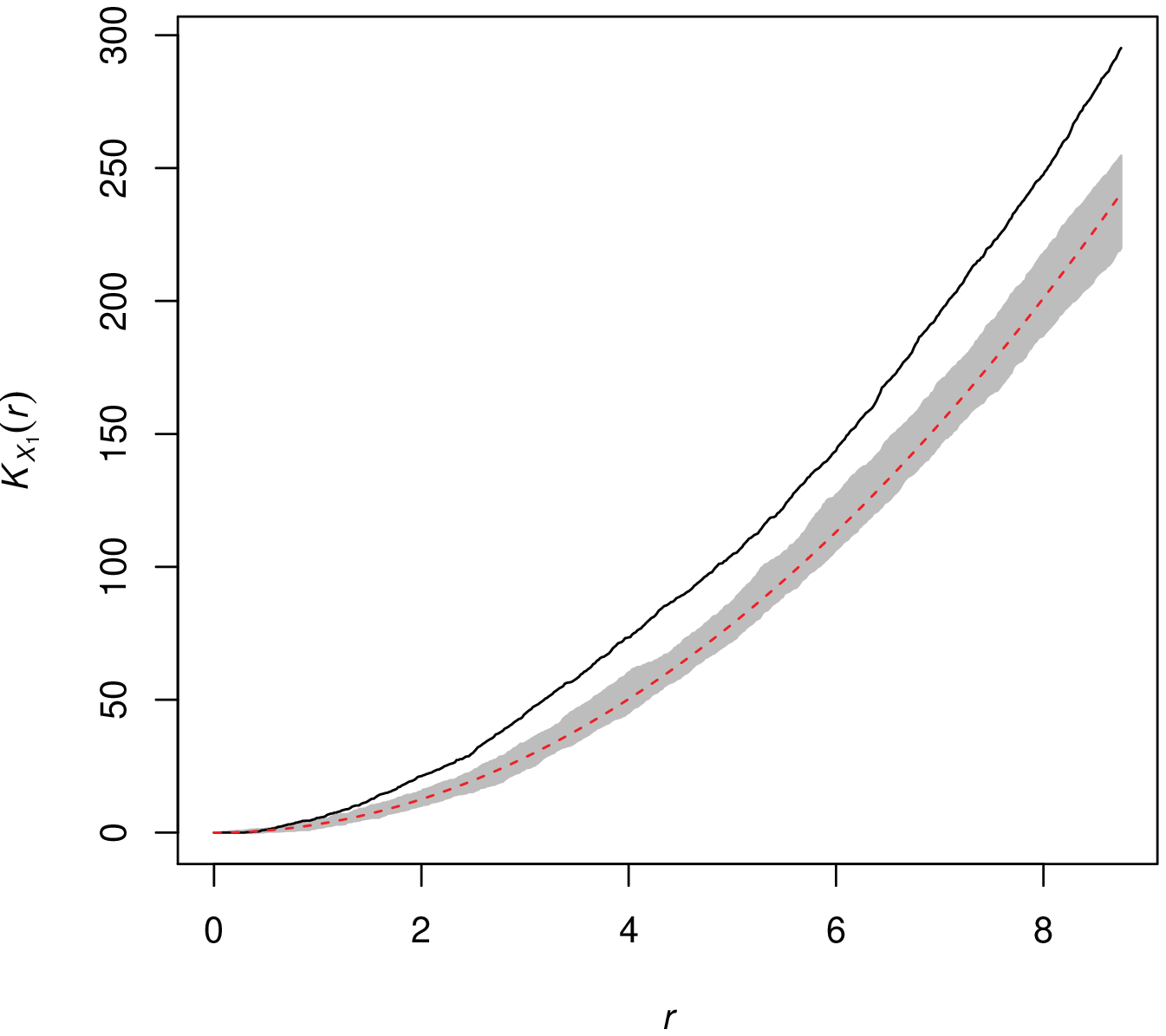}
        \caption{Sample $K$ function of $X_1$}\label{kX1}
    \end{subfigure}
   \begin{subfigure}{0.4\textwidth}
        \includegraphics[scale=0.3]{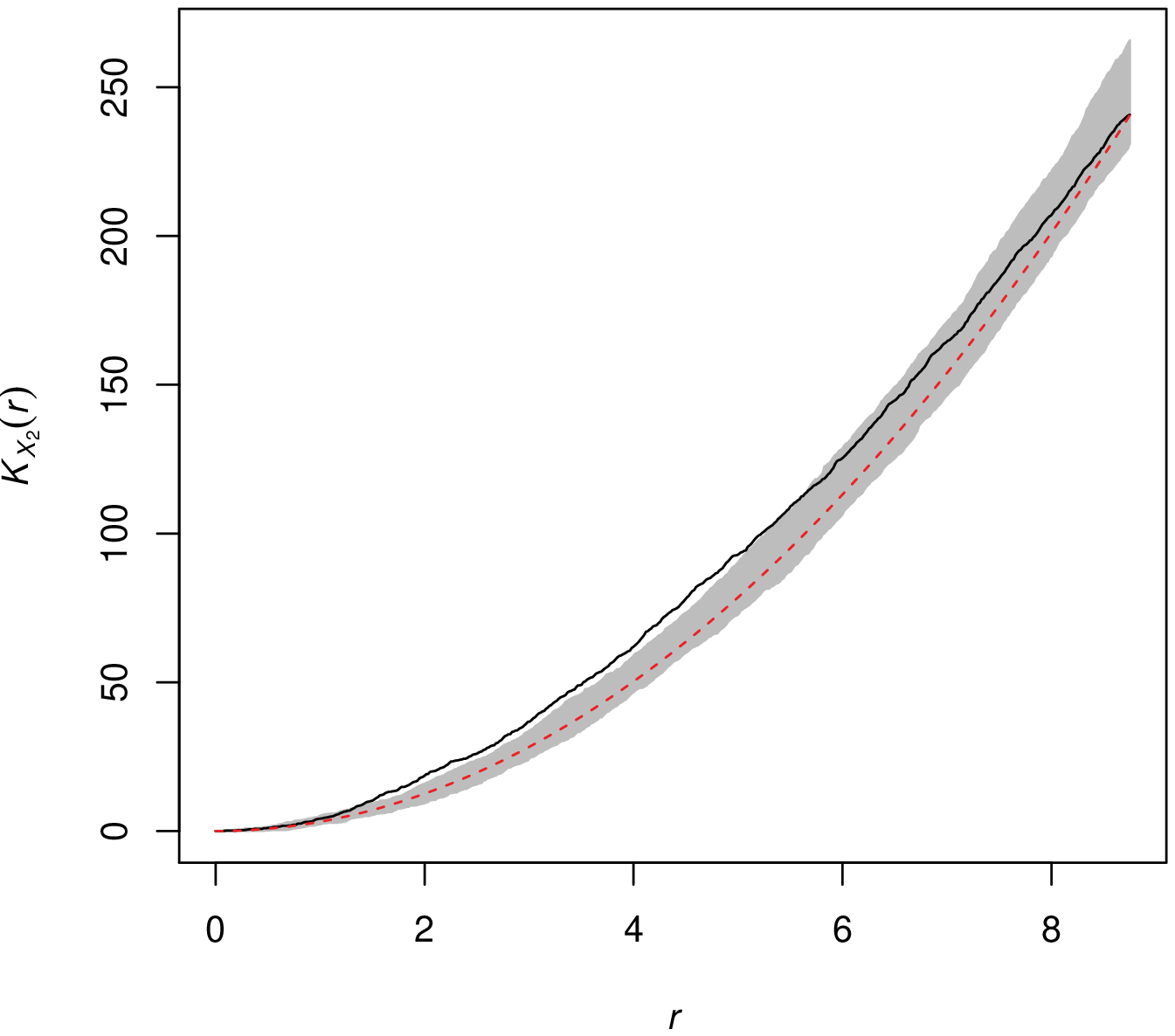}
        \caption{Sample $K$ function of $X_2$}\label{kX1}
    \end{subfigure}\\
    \begin{subfigure}{0.4\textwidth}
        \includegraphics[scale=0.3]{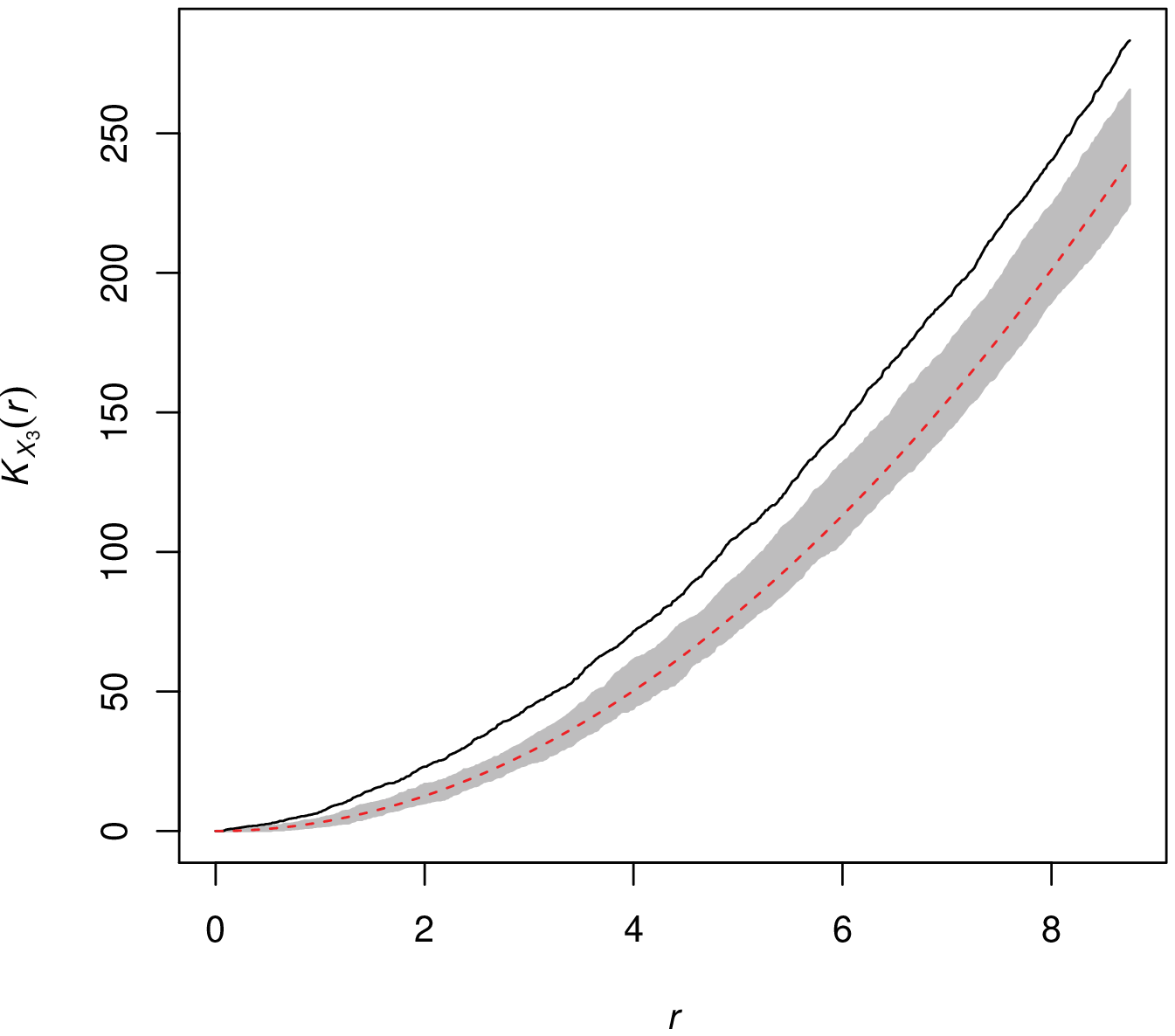}
        \caption{Sample $K$ function of $X_3$}\label{kX1}
    \end{subfigure}
    \begin{subfigure}{0.4\textwidth}
        \includegraphics[scale=0.3]{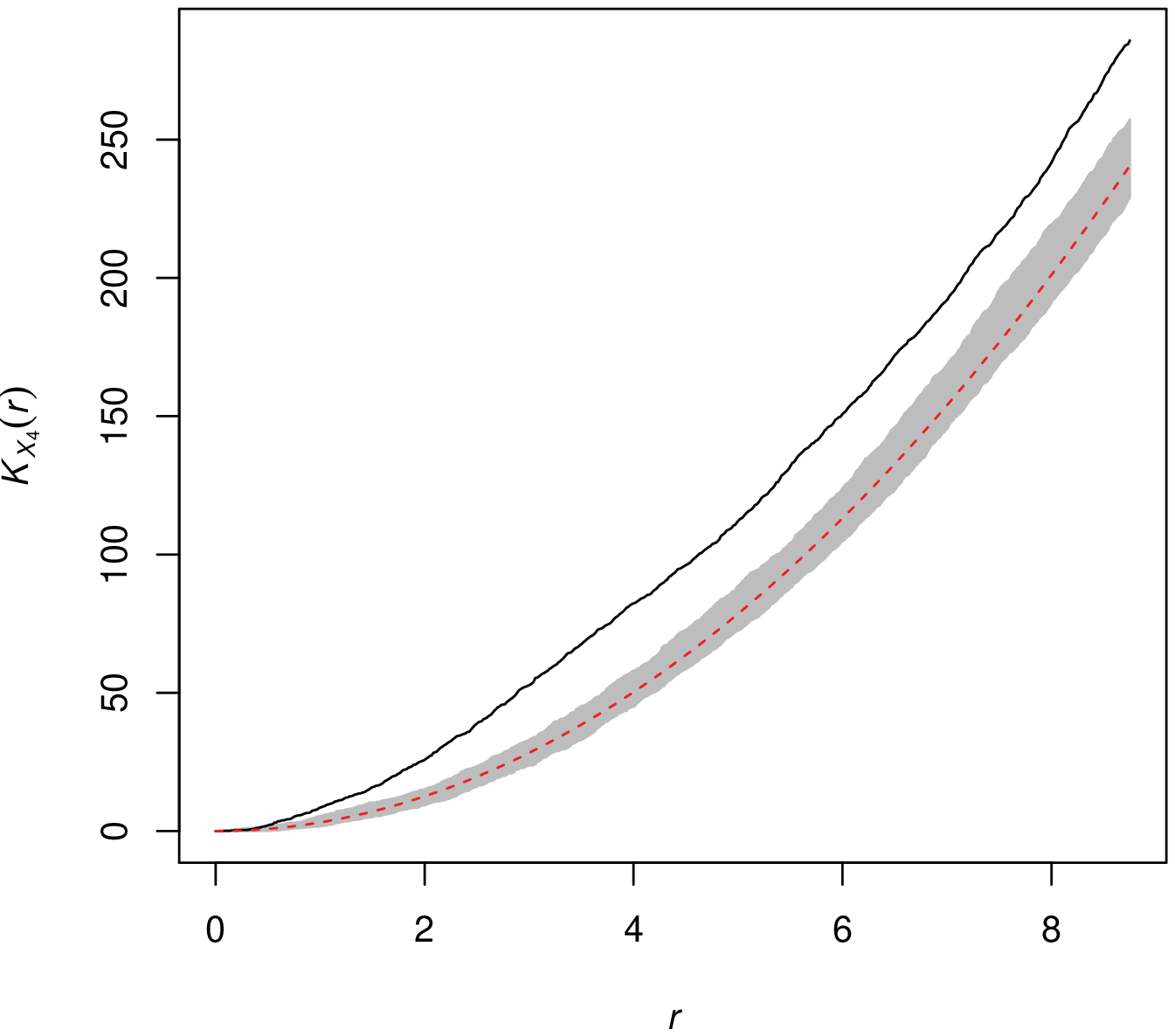}
        \caption{Sample $K$ function of $X_4$}\label{kX1}
    \end{subfigure}
\caption{The estimated $K$ function for trees data computed for point patterns restricted in 
square windows with centers $z_1,\ldots,z_4$ and the resulting point process is mentioned by 
$X_1,\ldots,X_4$, respectively. The figures contain the sample $K$ functions (solid line), the theoretic functions for the
stationary Poisson point process (dashed red line), and upper and lower
boundaries based on the enveloping of 99 simulations from stationary Poisson 
process (gray area).}\label{K function}
\end{figure}

As mentioned earlier, the first dataset is devoted to the locations of \textit{Euphorbiaceae} trees, as shown in Figure 
\ref{real data1}. Postulating that all of the assumptions made in Section \ref{simulation} are valid for this case, we consider 
$h=3$  and $\rho=34$. Thus, the distance between the locations $z_{i}$ and frequencies $\boldsymbol{\omega}_{j}$ 
points must be at least $\rho=34$ and $\pi/h=\pi/3$, respectively, and the value of $\sigma^{2}$ will be equal to 
$\sigma^{2}=16h^{2}/(9\rho^{2})=0.0138$. The points $ z_{1},\ldots,z_{4} $ are the centroids of the four equally-
dimensioned subregions denoted by $ \mathbf{S}_{1},\ldots,\mathbf{S}_{4} $ from left to right and down to up, 
respectively. In other words, the points $ z_{i}$ are chosen as $ z_{1}=(70/4,70/4)$, $ z_{2}=(3\times70/4,70/4)$, $ 
z_{3}=(70/4,3\times70/4)$, and $ z_{4}=(3\times70/4,3\times70/4)$. We denote the restricted point patterns to $ 
\mathbf{S}_{1},\ldots,\mathbf{S}_{4} $ by $ \mathbf{x}_{1},\ldots,\mathbf{x}_{4} $, respectively. The number of 
observed trees at the regions $ \mathbf{S}_{1},\ldots,\mathbf{S}_{4} $ are almost the same. Firstly, we consider the 
estimate of $K$ function for investigating the inter-point dependence aspects of these point patterns. The sample $K$ 
functions of each point pattern along the theoretic $K$ function of the stationary Poisson process are shown in Figure 
\ref{K function}. Simply speaking, the gray areas of all the figures represent the acceptance regions of the 
stationary Poisson 
model for data. The hypothesis of stationary Poisson model is rejected based on a statistic at level of $\alpha = 0.05$, 
when 
the sample statistic does not remain between the given boundaries at least for a point $ r \in \mathbb{R}_+ $. When the 
sample $K$ functions compared with the prepared boundaries, the lack of fitness of stationary Poisson process to all of the 
point patterns is emphasized. Visually, the sample $K$ functions of all the point patterns expect $ \mathbf{x}_{2}$ are 
approximately the same. In the following, we apply our proposed method to test the nonstationarity of this point pattern. 
Afterwards, if the test rejects the stationarity assumption against the zonal stationarity, then we use the post-hoc tests for 
detecting the location(s) at which the structure of trees pattern is changed. 

Table \ref{tab3} shows the results of post-hoc test. There is a significant difference between the local 
periodograms at locations $z_{2}$ and $z_{4}$, while the behavior of the local periodograms dose not change at other 
locations. Since the number of observed trees in both areas ($ \mathbf{S}_{2}$ and $ \mathbf{S}_{4}$) is almost the 
same, so the difference between the local periodograms at locations $z_{2}$ and $z_{4}$ may be due to the significance of 
the latitude effect on the competition between trees.
\begin{table}[h]
\caption{\small The Bonferroni post-hoc test for the trees data.}
\label{tab3}
\vspace{.25cm}
\centering
\footnotesize
\begin{tabular}{lcccc}
\hline
Item& Df&$\chi ^{2}(=SS/\sigma^2)$ statistic& $p$-value\\ 
\hline
$z_{1}$ vs $z_{2}$&1&1.38&$0.24$\\
$z_{1}$ vs $z_{3}$&1&0.12&$0.73$\\
$z_{1}$ vs $z_{4}$&1&4.99&$0.03$\\
$z_{2}$ vs $z_{3}$&1& 0.69&$0.41$\\
$z_{2}$ vs $z_{4}$&1&11.64&$<0.05/6=0.008$*\\
$z_{3}$ vs $z_{4}$&1&6.66&$0.01$\\
\hline                                                                                         
\end{tabular}
\end{table}

Table \ref{tab2} presents the results of the 
analysis of variance for this setting of locations and frequencies using the logarithm of local periodogram. The interaction 
term is not significant ($\chi^{2}$ is small compared to $\chi^{2}_{24}(0.05)= 36.42 $) confirming that the point pattern 
is uniformly modulated. Morever,  both `between spatial locations' and `between frequencies' sums of squares are highly 
significant ($SSL/\sigma^2 >\chi^{2}_{3}(0.05)=7.81 $ and $SSF/\sigma^2>\chi^{2}_{8}(0.05)=15.51$), suggesting 
that the point pattern is nonstationary and that the spectra are nonuniform. The analysis of variance indicates a significant 
difference in the locations effect. Thus, we use the Bonferroni method for multiple comparisons to discover different 
locations. There is a set of $ \binom{4}{2}=6 $ hypotheses to test, say, $ \alpha_{i}=\alpha_{j} $ for $ i\neq j $ and $ 
i,j=1,\ldots,4 $. The Bonferroni method rejects each test if $ SSL/ \sigma^{2} > \chi^{2}_{1}(\alpha/6) $. The Bonferroni 
method simply reduces the significance level of each individual test so that the sum of the significance levels is no greater 
than $ \alpha $. 
\begin{table}[h]
\caption{\small Analysis of variance table for the trees data.}
\label{tab2}
\vspace{.25cm}
\centering
\footnotesize
\begin{tabular}{lcccc}
\hline
Item& Df&$SS$&$\chi ^{2}(=SS/\sigma^2)$\\
\hline
Between spatial locations&3&0.18&12.74\\
Between frequencies&8&11.60&838.00\\
Interaction + residual&24&0.37&26.60\\
Total&35& 12.14&877.34\\
\hline                                                                                         
\end{tabular}
\end{table}

\subsubsection{Capillaries data}
The stationary Strauss hard-core model was suggested for the locations of capillaries in prostate 
tissues \cite{Mattfeldt2006, Mattfeldt2007}. It has been concluded that capillary profile patterns are more clustered in healthy tissue than in cancerous tissue the difference in the spatial model of healthy and cancerous tissues 
was verified \cite{Hahn2012} by testing the corresponding 
empirical $K$ functions. The intensities of two point patterns are almost the same. According to the stationarity assumption 
considered by \cite{Mattfeldt2006, Mattfeldt2007} to these patterns, the second order properties of 
healthy and cancerous tissues were compared \cite{saadat} and concluded that cancer does not affect the first and 
second order properties of the 
locations of capillaries on the prostate tissue. All these researches assumed the corresponding point process to be 
stationary. Here, we apply our proposed method to test the nonstationarity of both point 
patterns. The observation windows are rescaled to $ \left[ 0,70\right] ^{2}$ cubes and all the required 
values are assumed to be same as the the previous settings. 
\begin{table}[h]
\caption{\small Analysis of variance table for the Capillaries data.}
\label{tab4}
\vspace{.25cm}
\centering
\footnotesize
\begin{tabular}{lccc}
\hline
\multicolumn{4}{c}{Healthy tissue}\\
\hline 
source of variation& $df$& $SS$ &$\chi ^{2}$(=SS/$\sigma^{2}$)\\
\hline
Between spatial locations&3	&	0.21&15.00\\  
Between frequencies&8&2.77&200.03\\  
Interaction + residual&24&0.39	&27.84\\  
Total&35	&3.36&242.87\\  
\hline
\multicolumn{4}{c}{Cancerous tissue}\\
\hline 
source of variation& $df$& $SS$ &$\chi ^{2}$(=SS/$\sigma^{2}$)\\
\hline
Between spatial locations&3&	0.53&38.20\\  
Between frequencies&8	&	1.90&136.94\\  
Interaction + residual& 24	&0.25&17.96\\  
Total&35&2.67&193.09\\  
\hline
\end{tabular}
\end{table}

\begin{table}[h]
\caption{\small The Bonferroni post-hoc test for the Capillaries data. The significant 
differences in the location effects are denoted by `*'.}
\label{tab5}
\vspace{.25cm}
\centering
\footnotesize
\begin{tabular}{lccc}
\hline
\multicolumn{4}{c}{Healthy tissue}\\
\hline 
test& $df$&$\chi^{2}=SS/\sigma^{2}$& $p$-value\\	
\hline
$z_{1}$ vs $z_{2}$&1	&4.65	&$0.03$\\
$z_{1}$ vs $z_{3}$&1	&13.94	&$<0.05/6=0.008$*\\
$z_{1}$ vs $z_{4}$&1	&7.60	&$0.006$*	\\
$z_{2}$ vs $z_{3}$&1	&2.49	&$0.11$\\
$z_{2}$ vs $z_{4}$&1	&0.36	&$0.55$\\			
$z_{3}$ vs $z_{4}$&1	&0.95	&$0.33$\\			
\hline
\multicolumn{4}{c}{Cancerous tissue}\\
\hline 
test& $df$&$\chi^{2}=SS/\sigma^{2}$& $p$-value\\
\hline
$z_{1}$ vs $z_{2}$&1	&18.53	&$<0.05/6=0.008$*\\
$z_{1}$ vs $z_{3}$&1	&32.45	&$<0.008$*\\
$z_{1}$ vs $z_{4}$&1	&22.35	&$<0.008$*\\
$z_{2}$ vs $z_{3}$&1	&1.94&$0.16$\\
$z_{2}$ vs $z_{4}$&1	&0.18&$0.67$\\
$z_{3}$ vs $z_{4}$&1	&0.94&$0.33$\\
\hline
\end{tabular}
\end{table}

The results of the analysis of variance and post-hoc test for this 
dataset are presented at Table \ref{tab4} and \ref{tab5}, respectively. The results show that the location effect is significant for 
both of the healthy and cancerous point patterns. According to the new obtained evidence, the previous results could 
not be 
invoked. Nevertheless, we can use the local periodograms to extend the idea of \cite{saadat} for comparing the 
spectral density functions of nonstationary point patterns. The asymptotic independence 
and the asymptotic distribution of local periodograms are used to compute the density function of the local periodograms 
and hence the likelihood function in terms of the local periodograms similar to \cite{saadat}. Let $ I_{z_{i}}^{\mathbf{h}}
(\boldsymbol{\omega}_{j}) $ denotes to the local periodogram of healthy point pattern at the location $ z_{i} $ and 
frequency $ \boldsymbol{\omega}_{j} $ and there is similar notation, i.e., $ I_{z_{i}}^{\mathbf{c}}
(\boldsymbol{\omega}_{j}) $, for cancerous one. Thus, the likelihood ratio for comparing the local spectral density 
functions of two independent point patterns at the location $ z_{i} $  will be as below.
\begin{eqnarray*}
\Lambda_{ z_{i}}=\prod_{j=1}^{9}\dfrac{4I_{z_{i}}^{\mathbf{h}}(\boldsymbol{\omega}_{j}) I_{z_{i}}^{\mathbf{c}}
(\boldsymbol{\omega}_{j}) }{(I_{z_{i}}^{\mathbf{h}}(\boldsymbol{\omega}_{j}) +I_{z_{i}}^{\mathbf{c}}
(\boldsymbol{\omega}_{j}) )^{2}}
\end{eqnarray*}
Finally, the resulting likelihood ratio is compared with $ 0.025 $ and $ 0.975 $ quintiles estimated using the simple 
Monte Carlo simulation. The values of $ \Lambda_{ z_{i}}, i=1,...,4 $ are $ 0.93, 0.95, 0.89 $ and $ 0.94 $, 
respectively, and the estimated quintiles are $ 1.07\times10^{-5} $ and $0.20 $. The results of likelihood ratio test show that there 
are significant difference between the local spectral density functions of healthy and cancerous point patterns at the 
locations 
$ z_{i}, i=1,...,4 $. Therefore, assuming the nonstationarity of both point patterns, we can conclude that cancer affects the 
second order structure of prostate tissue.

%
%

%




\end{document}